\documentclass[a4paper,12pt]{article}

\usepackage{graphics, epsfig}

\begin{document}

\author{P. A. Hogan\thanks{E-mail : phogan@ollamh.ucd.ie} and E. M. O'Shea
\thanks{E-mail : emer.oshea.2@student.ucd.ie}\\
\small Mathematical Physics Department\\
\small  National University of Ireland Dublin, Belfield, Dublin 4, Ireland}

\title{Gravitational Wave Propagation in Isotropic Cosmologies}
\date{}
\maketitle

\begin{abstract}
We study the propagation of gravitational waves carrying arbitrary information 
through isotropic cosmologies. The waves are modelled as small perturbations 
of the background Robertson--Walker geometry. The perfect fluid matter distribution 
of the isotropic background is, in general, modified by small anisotropic stresses.
For pure gravity waves, in which the perturbed Weyl tensor is radiative (i.e. 
type N in the Petrov classification), we construct explicit examples for which 
the presence of the anisotropic stress 
is shown to be essential and the histories of the wave--fronts in the background 
Robertson--Walker geometry are shear--free null hypersurfaces. The examples 
derived in this case are analogous to the Bateman waves of 
electromagnetic theory.

\end{abstract}
\vskip 2truepc\noindent
PACS number(s): 04.30.Nk
\thispagestyle{empty}
\newpage

\section{Introduction}\indent

This paper is primarily concerned with investigating the influence of gravitational 
waves on the matter content of the universe. The gravitational waves are modelled as 
small perturbations of Friedmann--Lema\^itre cosmological models with a spatially 
homogeneous and isotropic Robertson--Walker (RW) geometry. We use the gauge--
invariant and covariant perturbation approach of Ellis and Bruni \cite{EB}. This 
technique has previously been used to study aspects of gravitational wave propagation 
in isotropic cosmologies which differ from those considered here (see, for example, 
\cite{DBE}, \cite{MDB}).

We look for gravitational waves which carry arbitrary information and this 
motivates us to require the Ellis--Bruni gauge--invariant variables to have 
an arbitrary dependence on a function. Thus by ``arbitrary information" in 
the perturbations we mean that they depend upon an arbitrary function. This in 
turn means that the profile of the waves described by the perturbations is 
unspecified. The idea of introducing arbitrary functions into solutions of 
Einstein's equations 
describing gravitational waves goes back to pioneering work by Trautman \cite{T}.
This point of view was initiated in the present context by 
Hogan and Ellis \cite{HE}, but whereas in 
that work the perturbed matter distribution was a perfect fluid we here allow it 
to be completely general. The Ellis--Bruni approach does not involve working 
directly with Einstein's field equations but instead the basic equations used 
are the Ricci identities, Bianchi identities and the matter equations of motion 
and energy conservation equation, with Einstein's field equations incorporated 
in them. We work in the linear approximation in terms of the Ellis--Bruni variables 
and demonstrate that consistency of our assumptions with the basic equations necessarily 
leads to all perturbation variables vanishing except the perturbed shear of 
the matter world--lines and the anisotropic stress perturbation of the matter 
distribution. The consistency of the equations satisfied by these 
surviving variables is established. We then specialise our study to pure gravity 
wave perturbations for which the perturbed Weyl tensor is radiative (type N in 
the Petrov classification). We show that for such waves the histories of their 
wave--fronts in the background RW space--time can be shear--free 
null hypersurfaces. The various open and closed RW geometries admit, in 
a natural way, families of shear--free null hypersurfaces. We derive explicit 
solutions of the perturbation equations describing gravitational waves 
propagating through these isotropic universes for which the histories of 
their wave fronts are these naturally occuring families of shear--free null 
hypersurfaces. These solutions are analogous to the Bateman waves \cite{B} 
in electromagnetic theory.

The paper is organised as follows: in section 2 we list the basic exact equations 
derived from the Ricci identities, Bianchi identities and the equations of motion 
and of energy conservation of the matter distribution. In section 3 we 
introduce the Ellis--Bruni gauge--invariant variables and show how for us they 
have an arbitrary dependence on a function. The consistency of these assumptions 
with the basic equations in the linear approximation is then systematically 
studied, and the mathematical consistency of the surviving equations is 
established. This is followed in section 4 by specialisation to pure gravity 
wave perturbations and the demonstration that their wave--front histories can 
be shear--free null hypersurfaces. In section 5 some explicit families of pure 
gravity wave perturbations are derived. The paper ends with a discussion in 
which we comment on the solutions obtained in section 5.

\setcounter{equation}{0}
\section{The Basic Exact Equations}\indent

To make the paper as self--contained as possible the basic equations 
required for our study are given in this section. We use the notation and sign conventions of 
\cite{Ellis} and all of the equations given here can 
be found in \cite{Ellis} with the exception that the Bianchi identities 
given in \cite{Ellis} apply to a perfect fluid matter distribution 
whereas we require the extension of these to a general matter distribution 
with energy--momentum--stress tensor given by (\ref{2.2}) below. This 
covariant approach to cosmology began in a systematic way with the work 
of Sch\"ucking, Ehlers and Sachs [see \cite {Ellis}, \cite{Ehlers} 
for example] while Hawking 
\cite{Hawking} gave the first description of cosmological perturbations in 
this context. We are concerned here with a four dimensional space--time manifold with 
metric tensor components $g_{ab}$, in a local coordinate system $\{x^a\}$, 
and a preferred congruence of world--lines tangent to the unit 
time--like vector field with components $u^a$ (with $u^a\,u_a=-1$). 
With respect to this 4--velocity field the Weyl tensor, with components 
$C_{abcd}$, is decomposed into an ``electric part" and a ``magnetic 
part" given respectively by
\begin{equation}\label{2.1}
E_{ab}=C_{apbq}\,u^p\,u^q\ ,\qquad H_{ab}={}^*C_{apbq}\,u^p\,u^q\ .
\end{equation}
Here ${}^*C_{apbq}=\frac{1}{2}\eta_{ap}{}^{rs}\,C_{rsbq}$ is the dual of 
the Weyl tensor (the left and right duals being equal), $\eta _{abcd}
=\sqrt{-g}\,\epsilon _{abcd}$ with $g={\rm det}\,(g_{ab})$ and 
$\epsilon _{abcd}$ is the Levi--Civita permutation symbol. The expression 
for the Weyl tensor in terms of $E_{ab}$ and $H_{ab}$ is given in \cite{Ellis}. The symmetric 
energy--momentum--stress tensor $T^{ab}$ can be decomposed with respect to the 
4--velocity field $u^a$ as
\begin{equation}\label{2.2}
T^{ab}=\mu\,u^a\,u^b+p\,h^{ab}+q^a\,u^b+q^b\,u^a+\pi ^{ab}\ ,
\end{equation}
where
\begin{equation}\label{2.3}
h^{ab}=g^{ab}+u^a\,u^b\ ,
\end{equation}
is the projection tensor and
\begin{equation}\label{2.4}
q^a\,u_a=0\ ,\qquad \pi ^{ab}\,u_b=0\ ,\qquad \pi ^a{}_a=0\ ,
\end{equation}
with $\pi ^{ab}=\pi ^{ba}$. Here $\mu$ is interpreted as the matter energy 
density measured by the observer with 4--velocity $u^a$, $p$ is the 
isotropic pressure, $q^a$ is the energy flow (such as heat flow) 
measured by this observer and $\pi ^{ab}$ is the anisotropic stress 
(due, for example, to viscosity). We shall indicate covariant differentiation 
with a semi--colon, partial differentiation by a comma, covariant differentiation in the direction of 
$u^a$ by a dot and a definition by a colon followed by an equality sign. 
Thus the 4--acceleration of the time--like congruence is
\begin{equation}\label{2.5}
\dot u^a:=u^a{}_{;b}\,u^b\ ,
\end{equation}
and $u_{a;b}$ can be decomposed into
\begin{equation}\label{2.6}
u_{a;b}=\omega _{ab}+\sigma _{ab}+\frac{1}{3}\theta\,h_{ab}-\dot u_a\,u_b\ ,
\end{equation}
with
\begin{equation}\label{2.7}
\omega _{ab}:=u_{[a;b]}+\dot u_{[a}\,u_{b]}\ ,
\end{equation}
the vorticity tensor of the congruence tangent to $u^a$. The square brackets 
denote skew--symmetrisation as usual. Also
\begin{equation}\label{2.8}
\sigma _{ab}:=u_{(a;b)}+\dot u_{(a}\,u_{b)}-\frac{1}{3}\theta\,h_{ab}\ ,
\end{equation}
is the shear tensor of the congruence. The round brackets denote symmetrisation and 
\begin{equation}\label{2.9}
\theta :=u^a{}_{;a}\ ,\end{equation}
is the expansion (or contraction) of the congruence. 

The key equations we shall need are obtained by projections in the direction 
$u^a$ and orthogonal to $u^a$ (using the projection tensor $h_{ab}$) of the 
{\it Ricci identities}
\begin{equation}\label{2.10}
u_{a;dc}-u_{a;cd}=R_{abcd}\,u^b\ ,
\end{equation}
where $R_{abcd}$ is the Riemann curvature tensor, the {\it equations 
of motion} and the {\it energy conservation equation} contained in 
\begin{equation}\label{2.11}
T^{ab}{}_{;b}=0\ ,
\end{equation}
and the {\it Bianchi identities} written conveniently in the form
\begin{equation}\label{2.12}
C^{abcd}{}_{;d}=R^{c[a;b]}-\frac{1}{6}\,g^{c[a}\,R^{;b]}\ .
\end{equation}
Here $R^{ca}:=R^{cba}{}_{b}$ are the components of the Ricci tensor and 
$R:=R^c{}_{c}$ is the Ricci scalar. Einstein's field equations, after 
absorbing the coupling constant into the energy--momentum--stress tensor, 
take the form
\begin{equation}\label{2.13}
R_{ab}-\frac{1}{2}\,g_{ab}\,R=T_{ab}\ .
\end{equation}

The Ricci identities yield {\it Raychaudhuri's equation},
\begin{equation}\label{2.14}
\dot\theta +\frac{1}{3}\,\theta ^2-\dot u^a_{;a}+2\,(\sigma ^2-\omega ^2)+
\frac{1}{2}\,(\mu +3\,p)=0\ ,
\end{equation}
(here $\sigma ^2:=\frac{1}{2}\sigma _{ab}\,\sigma ^{ab}\ ,\ 
\omega ^2:=\frac{1}{2}\omega _{ab}\,\omega ^{ab}\ $) the {\it vorticity 
propagation equation},
\begin{equation}\label{2.15}
h^a_b\,\dot\omega ^b+\frac{2}{3}\,\theta\,\omega ^a=\sigma ^a{}_b\,\omega ^b+\frac{1}{2}\eta ^
{abcd}\,u_b\,\dot u_{c;d}\ ,
\end{equation}
(here $\omega ^a:=\frac{1}{2}\,\eta ^{abcd}\,u_b\,\omega _{cd}$ is the 
vorticity vector) the {\it shear propagation equation},
\begin{eqnarray}\label{2.16}
&{}&h^f_a\,h^g_b\,(\dot\sigma _{fg}-\dot u_{(f;g)})-\dot u_a\,\dot u_b+
\omega _a\,\omega _b+\sigma _{af}\,\sigma ^f{}_b+
\frac{2}{3}\,\theta\,\sigma _{ab}\nonumber\\
&{}&+h_{ab}\,\left (-\frac{1}{3}\omega ^2-\frac{2}{3}\sigma ^2+\frac{1}
{3}\dot u^c_{;c}\right )-\frac{1}{2}\pi _{ab}+E_{ab}=0\ ,
\end{eqnarray}
the $(0, \nu )$--{\it field equation} (see \cite{Ellis} for the explanation of 
this terminology),
\begin{equation}\label{2.17}
\frac{2}{3}\,h^a_b\,\theta ^{,b}-h^a_b\,\sigma ^{bc}{}_{;d}\,h^d_c-
\eta ^{acdf}\,u_c\,(\omega _{d;f}+2\,\omega _d\,\dot u_f)=q^a\ ,
\end{equation}
the {\it divergence of vorticity equation},
\begin{equation}\label{2.18}
\omega ^a{}_{;b}\,h^b_a=\omega ^a\,\dot u_a\ ,
\end{equation}
and the {\it magnetic part of the Weyl tensor},
\begin{equation}\label{2.19}
H_{ab}=2\,\dot u_{(a}\,\omega _{b)}-h^t_a\,h^s_b\,\left (\omega _{(t}{}^{g;c}
+\sigma _{(t}{}^{g;c}\right )\,\eta _{s)fgc}\,u^f\ .
\end{equation}

Next the equations (\ref{2.11}) projected orthogonal to $u^a$ and along $u^a$ respectively 
give the {\it equations of motion of matter},
\begin{equation}\label{2.20}
(\mu +p)\,\dot u^a+h^{ac}\,(p_{,c}+\pi ^b{}_{c;b}+\dot q_c)+
(\omega ^{ab}+\sigma ^{ab}+\frac{4}{3}\,\theta\,h^{ab})\,q_b=0\ 
,\end{equation}
and the {\it energy conservation equation},
\begin{equation}\label{2.21}
\dot\mu +\theta\,(\mu +p)+\pi _{ab}\,\sigma ^{ab}+q^a{}_{;a}+\dot u^a\,q_a=0\ .
\end{equation}

Finally the various projections of the Bianchi identities (\ref{2.12}) along 
and orthogonal to $u^a$ give analogous equations to Maxwell's 
equations for $E_{ab}$ and $H_{ab}$ in (\ref{2.1}). These consist of 
the {\it div--$E$ equation},
\begin{eqnarray}\label{2.22}
&{}&h^b_g\,E^{gd}{}_{;f}\,h^f_d+3\,\omega ^s\,H^b_s-\eta ^{bapq}\,u_a\,\sigma ^d{}_p\,H_{qd}=
\frac{1}{3}\,h^b_c\,\mu ^{,c}\nonumber\\
&{}&+\frac{1}{2}\left\{-\pi ^{bd}{}_{;d}+u^b\,\sigma _{cd}\,\pi ^{cd}
-3\,\omega ^{bd}\,q_d+\sigma ^{bd}\,q_d-\frac{2}{3}\,\theta\,q^b+\pi ^{bd}\,\dot u_d
\right\}\ ,
\end{eqnarray}
the {\it div--$H$ equation},
\begin{eqnarray}\label{2.23}
&{}&h^b_g\,H^{gd}{}_{;f}\,h^f_d-3\,\omega ^s\,E^b_s+\eta ^{bapq}\,u_a\,\sigma ^d{}_p\,E_{qd}=
(\mu +p)\,\omega ^b\nonumber\\
&{}&+\frac{1}{2}\,\eta ^{b}{}_{qac}\,u^q\,q^{a;c}+\frac{1}{2}\,\eta ^{b}{}_{qac}\,u^q\,(\omega ^{dc}
+\sigma ^{dc})\,\pi ^{a}{}_{d}\ ,
\end{eqnarray}
the {\it $\dot E$--equation},
\begin{eqnarray}\label{2.24}
&{}&h^b_f\,\dot E^{fg}\,h^t_g+h^{(b}_a\,\eta ^{t)rsd}\,u_rH^a_{s;d}-
2\,H^{(b}_s\,\eta ^{t)drs}\,u_d\dot u_r\nonumber\\
&{}&-E^{(t}_s\,\omega ^{b)s}
-3\,E^{(t}_s\,\sigma ^{b)s}+h^{tb}\,E^{dp}\,\sigma _{dp}
+\theta\,E^{bt}=-\frac{1}{2}\,(\mu +p)\,\sigma ^{tb}\nonumber\\
&{}&-\frac{1}{6}\,h^{tb}\,\{\dot\mu +\theta\,(\mu +p)\}
-\,q^{(b}\,\dot u^{t)}-
\frac{1}{2}\,u^{(b}\,\dot q^{t)}-\frac{1}{2}\,q^{(t;b)}\nonumber\\
&{}&+\frac{1}{2}\{\omega ^{c(b}+\sigma ^{c(b}\}\,u^{t)}\,q_c+\frac{1}{6}\,\theta\,u^{(t}\,
q^{b)}-\frac{1}{2}\,\dot\pi +\pi ^{c(b}\,u^{t)}\,\dot u_c\nonumber\\
&{}&-\frac{1}{2}\,\{\omega ^{c(b}+\sigma ^{c(b}\}\,\pi ^{t)}{}_c-\frac{1}{6}\,\theta\,\pi 
^{bt}\ ,
\end{eqnarray}
and the {\it $\dot H$--equation}, 
\begin{eqnarray}\label{2.25}
&{}&h^b_f\,\dot H^{fg}\,h^t_g-h^{(b}_a\,\eta ^{t)rsd}\,u_rE^a_{s;d}+
2\,E^{(b}_s\,\eta ^{t)drs}\,u_d\dot u_r\nonumber\\
&{}&-H^{(t}_s\,\omega ^{b)s}
-3\,H^{(t}_s\,\sigma ^{b)s}+h^{tb}\,H^{dp}\,\sigma _{dp}
+\theta\,H^{bt}=-q^{(t}\,\omega ^{b)}\nonumber\\
&{}&-\frac{1}{2}\,\eta ^{(t}{}_{rad}\,\{\omega ^{b)d}+\sigma ^{b)d}\}\,u^r\,q^a-
\frac{1}{2}\eta ^{(b}{}_{rad}\,\pi ^{t)a;d}\,u_r\nonumber\\
&{}&+\frac{1}{2}\eta ^{(b}{}_{rad}\,u^{t)}\,u^r\,\{\omega ^{cd}+\sigma ^{cd}\}\,\pi ^a{}_c\ .
\end{eqnarray}

\setcounter{equation}{0}
\section{Perturbations of Isotropic Cosmologies}\indent

We shall now assume that the space--time in section 2 is a perturbed RW space--time. 
Thus the background metric tensor $g_{ab}$ is the Robertson--Walker metric, the 
background energy--momentum--stress tensor is (\ref{2.2}) specialised to 
a perfect--fluid (by putting $q^a=0=\pi ^{ab}$) with fluid 4--velocity $u^a$ and 
the background Weyl tensor vanishes. The Ellis--Bruni \cite{EB} approach to 
perturbations of this background is to work with gauge--invariant small quantities 
rather than small perturbations of the background metric. Such gauge--invariant 
quantities have the important property that they vanish in the background space--time. 
Thus for an isotropic background the Ellis--Bruni variables, which will henceforth be 
considered small of first order, are $E_{ab},\  H_{ab},\  \sigma _{ab},\  \dot u^a,\  
\omega _{ab}$ (or equivalently the vorticity vector $\omega ^a$), $X_a=h^b_a\,\mu _{,b}, 
\ Y_a=h^b_a\,p_{,b}\,,\  Z_a=h^b_a\,\theta _{,b}\,,\  \pi _{ab}$ and $q^a$. The equations 
satisfied by these quantities are given by (\ref{2.15})--(\ref{2.20}) and by 
(\ref{2.22})--(\ref{2.25}) neglecting non--linear terms in these variables. Raychaudhuri's 
equation (\ref{2.14}) and the energy conservation equation (\ref{2.21}) are not 
immediately useful because they are not expressed in terms of these gauge--invariant 
variables. We can however get useful equations from (\ref{2.14}) and (\ref{2.21}) 
by calculating the spatial gradients of these equations and then retaining only 
linear terms in the gauge--invariant variables. This results in the two equations
\begin{equation}\label{3.1}
\dot Z^a+\theta\,Z^a-\dot\theta\,\dot u^a-h^{ab}\,\left (\dot u^a{}_{;a}\right )_{,b}
+\frac{1}{2}\,X^a+\frac{3}{2}\,Y^a=0\ ,
\end{equation}
and 
\begin{equation}\label{3.2}
\dot X^a+\frac{4}{3}\,\theta\,X^a +(\mu +p)\,Z^a+\theta\,Y^a-\dot\mu\,\dot u^a+
h^{ab}\,\left (q^c{}_{;c}\right )_{,b}=0\ .
\end{equation} 
The background value of $\dot\theta$ to be substituted into (\ref{3.1}) is given by 
\begin{equation}\label{3.3}
\dot\theta =-\frac{1}{3}\,\theta ^2-\frac{1}{2}\,(\mu +3\,p)\ ,
\end{equation}
which is got by specialising Raychaudhuri's equation (\ref{2.14}) to the background. 
In (\ref{3.2}) the background value of $\dot\mu$ is given by 
\begin{equation}\label{3.4}
\dot\mu =-\theta\,(\mu +p)\ ,
\end{equation}
which follows from (\ref{2.21}) specialised to the background. We shall assume an 
equation of state of the form $p=p(\mu )$ so that the Ellis--Bruni variables 
$X_a$ and $Y_a$ are related by
\begin{equation}\label{3.5}
Y_a=\frac{dp}{d\mu}\,X_a\ .
\end{equation}
Finally we shall assume that in the background $\mu +p\neq 0$ so that we do 
indeed have a cosmological model in which the Einstein tensor determines a 
unique time--like 4--velocity of the matter.

We look for solutions $\sigma _{ab}\ ,\ \dot u^a\ ,\ \omega _{ab}\ (\Leftrightarrow \omega ^a)\ ,
\ X_a$ (and thus $Y_a$ by (\ref{3.5})), $Z^a\ ,\ \pi ^{ab}$ and $q^a$ of the linearised versions of the 
equations (\ref{2.15})--(\ref{2.20}) and (\ref{2.22})--(\ref{2.25}) along with (\ref{3.1}) 
and (\ref{3.2}) for which these variables depend upon an arbitrary function. This is because 
we expect that this dependence of perturbations will describe gravitational waves carrying 
arbitrary information. We note that $E_{ab}\ ,\ H_{ab}$ are derived variables. Specifically 
we assume that 
\begin{eqnarray}\label{3.6}
\sigma _{ab}&=&s_{ab}\,F(\phi )\ ,\ \dot u^b=a^b\,F(\phi )\ ,\ \omega ^{ab}=w^{ab}\,
F(\phi )\ (\Leftrightarrow\ \omega ^a=w^a\,F(\phi ))\ ,\nonumber\\
X^a&=&x^a\,F(\phi )\ ,\ Z^a=z^a\,F(\phi )\ ,\ \pi ^{ab}=\Pi ^{ab}\,F(\phi )\ ,\ 
q^a=Q^a\,F(\phi )\ .
\end{eqnarray} 
where $F$ is an arbitrary real--valued function of its argument $\phi (x^a)$. This 
form for the gauge--invariant variables was first introduced by Hogan and Ellis \cite{HE} 
where the perturbed matter distribution was taken to be a perfect fluid. As we mentioned 
in the introduction, the idea of introducing arbitrary functions into solutions of Einstein's equations 
describing gravitational waves goes back to pioneering work by 
Trautman \cite{T}. We note that all of the quantities in (\ref{3.6}) are 
orthogonal to $u^a$ and that $s_{ab}\ ,\ \Pi _{ab}$ are trace--free 
with respect to the background metric $g_{ab}$ (i.e. $s^a{}_a=0=\Pi ^a{}_a$). 

When (\ref{3.6}) are substituted into the linearised versions of (\ref{2.15})--
(\ref{2.20}), (\ref{2.22})--(\ref{2.25}) and (\ref{3.1}), (\ref{3.2}) the 
following extensive but surveyable list of equations emerges as indicated.
\vskip 1truepc
\noindent
{\it From the shear propagation equation:}
\begin{equation}\label{3.7}
E_{ab}=\left (\frac{1}{2}\,\Pi _{ab}+p_{ab}\right ) F+m_{ab}\,F'\ ,
\end{equation}
with $F'=dF/d\phi$ and
\begin{equation}\label{3.8}
p_{ab}=a_{(a;b)}+u_{(a}\,\dot a_{b)}-\frac{1}{3}\,\theta\,u_{(a}\,a_{b)}
-\frac{1}{3}\,a^f{}_{;f}\,h_{ab}-\dot s_{ab}-\frac{2}{3}\,\theta\,s_{ab}\ ,
\end{equation}
\begin{equation}\label{3.9}
m_{ab}=a_{(a}\,\lambda _{b)}-\frac{1}{3}\,\chi\,h_{ab}-\dot\phi\,s_{ab}\ .
\end{equation}
Here and throughout $\lambda _a=h^b_a\,\phi _{,b}\ ,\ \chi =\phi _{,f}
\,a^f$ and $\dot\phi =\phi _{,a}\,u^a$.

\vskip 1truepc
\noindent
{\it From the magnetic part of the Weyl tensor:}
\begin{equation}\label{3.10}
H_{ab}=q_{ab}\,F+l_{ab}\,F'\ ,
\end{equation}
with
\begin{equation}\label{3.11}
q_{ab}=w_{(a;b)}+\dot w_{(a}\,u_{b)}-\frac{1}{3}\,\theta\,w_{(a}\,u_{b)}
-s_{(a}{}^{p;c}\,\eta _{b)fpc}\,u^f-h_{ab}\,w^c{}_{;c}\ ,
\end{equation}
\begin{equation}\label{3.12}
l_{ab}=w_{(a}\,\lambda _{b)}-w_c\,\phi ^{,c}\,h_{ab}-s_{(a}{}^p\,\eta _{
b)fpc}\,u^f\,\phi ^{,c}\ .
\end{equation}
\vskip 1truepc
\noindent
{\it From the div--E equation:}
\begin{equation}\label{3.13}
m^{ab}\,\phi _{,b}=0\ ,
\end{equation}
\begin{equation}\label{3.14}
\Pi ^{ab}\,\phi _{,b}+p^{ab}\,\phi _{,b}+m^{ab}{}_{;b}=0\ ,
\end{equation}
\begin{equation}\label{3.15}
\Pi ^{ab}{}_{;b}+p^{ab}{}_{;b}=\frac{1}{3}\,x^a-\frac{1}{3}\,\theta\,Q^a\ .
\end{equation}
\vskip 1truepc
\noindent
{\it From the div--H equation:}
\begin{equation}\label{3.16}
l^{ab}\,\phi _{,b}=0\ ,
\end{equation}
\begin{equation}\label{3.17}
q^{ab}\,\phi _{,b}+l^{ab}{}_{;b}=\frac{1}{2}\,\eta ^a{}_{qbc}\,u^q\,Q^b\,\phi ^{,c}
\ ,
\end{equation}
\begin{equation}\label{3.18}
q^{ab}{}_{;b}-(\mu +p)\,w^a=\frac{1}{2}\,\eta ^a{}_{qbc}\,u^q\,Q^{b;c}\ .
\end{equation} 
\vskip 1truepc
\noindent
{\it From the $\dot E$--equation:}
\begin{eqnarray}\label{3.19}
&&\dot \Pi^{bt}+\frac{2}{3}\,\theta\,\Pi ^{bt}-\frac{1}{6}\,h^{bt}\,Q^a{}_{;a}
+\frac{1}{2}\,u^{(t}\,\dot Q^{b)}+
\frac{1}{2}\,Q^{(t;b)}-\frac{1}{6}\,\theta
\,u^{(t}\,Q^{b)}=\nonumber\\
&&-\dot p^{bt}-u_r\,q^{(b}{}_{s;d}\,\eta ^{t)rsd}
-\theta\,p^{bt}-\frac{1}{2}\,(\mu +p)\,s^{bt}\ ,
\end{eqnarray}  
\begin{equation}\label{3.20}
\dot\phi\,\Pi  ^{bt}-\frac{1}{6}\,h^{bt}\,Q^a\,\phi _{,a}
+\frac{1}{2}\,Q^{(t}\,\lambda ^{b)}=-\dot\phi\,p^{bt}-\dot m^{bt}-\theta\,m^{bt}-u_r\,(q^{(b}{}_s\,\phi _{,d}+l^{(b}{}_{s;d}
)\,\eta ^{t)rsd}\ ,
\end{equation}
\begin{equation}\label{3.21}
\dot\phi\,m^{bt}+l^{(b}{}_s\,\eta ^{t)rsd}\,u_r\,\phi _{,d}=0\ .
\end{equation}
\vskip 1truepc
\noindent
{\it From the $\dot H$--equation:}
\begin{equation}\label{3.22}
\dot q^{bt}-u_r\,p^{(b}{}_{s;d}\,\eta ^{t)rsd}+\theta\,q^{bt}=0\ ,
\end{equation}
\begin{equation}\label{3.23}
\dot\phi\,q^{bt}+\dot l^{bt}+\theta\,l^{bt}-u_r\,(p^{(b}{}_s\,\phi _{,d}+
m^{(b}{}_{s;d})\,\eta ^{t)rsd}=0\ ,
\end{equation}
\begin{equation}\label{3.24}
\dot\phi\,l^{bt}-m^{(b}{}_s\,\eta ^{t)rsd}\,u_r\,\phi _{,d}=0\ .
\end{equation}
\vskip 1truepc
\noindent
{\it From the $(0, \nu )$--field equation:}
\begin{equation}\label{3.25}
s^{ab}\,\phi _{,b}+\eta ^{acdf}\,u_c\,w_d\,\phi _{,f}=0\ ,
\end{equation}
\begin{equation}\label{3.26}
Q^a=\frac{2}{3}z^a-s^{ab}{}_{;b}-
\eta ^{acdf}\,u_c\,w_{d;f}\ .
\end{equation}
\vskip 1truepc
\noindent
{\it From the vorticity propagation equation:}
\begin{equation}\label{3.27}
\dot w^a+\frac{2}{3}\,\theta\,w^a=\frac{1}{2}\,\eta ^{acbd}u_c\,a_{b;d}\ ,
\end{equation}
\begin{equation}\label{3.28}
\dot\phi\,w^a=\frac{1}{2}\,\eta ^{acbd}\,u_c\,a_b\,\phi _{,d}\ .
\end{equation}
\vskip 1truepc
\noindent
{\it From the divergence of vorticity equation:} 
\begin{equation}\label{3.29}
w^a{}_{;a}=0\ ,
\end{equation}
\begin{equation}\label{3.30}
w^a\,\phi _{,a}=0\ .
\end{equation}
\vskip 1truepc
\noindent
{\it From the spatial gradient of Raychaudhuri's equation:}
\begin{equation}\label{3.31}
\dot z^c+\theta\,z^c+\frac{1}{2}\,x^c+\frac{3}{2}\,y^c=
h^{cb}\,\left (a^d{}_{;d}\right )_{,b}+\dot\theta\,a^c\ ,
\end{equation}
\begin{equation}\label{3.32}
\dot\phi\,z^c=a^d{}_{;d}\,\lambda ^c+h^{cb}\,\left (a^d\,\phi _{,d}\right )_{,b}\ ,
\end{equation}
\begin{equation}\label{3.33}
\left (a^d\,\phi _{,d}\right )\,\lambda _c=0\ .
\end{equation}
\vskip 1truepc
\noindent
{\it From the equations of motion of matter:}
\begin{equation}\label{3.34}
y^a+\dot Q^a=-(\mu +p)\,a^a-\Pi ^{ab}{}_{;b}\ ,
\end{equation}
\begin{equation}\label{3.35}
\dot\phi\,Q^a=-\Pi ^{ab}\,\phi _{,b}\ .
\end{equation}
\vskip 1truepc
\noindent
{\it From the spatial gradient of the energy conservation equation:}
\begin{equation}\label{3.36}
\dot x^c+\frac{4}{3}\,\theta\,x^c+\theta\,y^c+h^{cb}\,\left (Q^a{}_{;a}
\right )_{,b}=-(\mu +p)\,\theta\,a^c-(\mu +p)\,z^c\ ,
\end{equation}
\begin{equation}\label{3.37}
\dot\phi\,x_c+Q^a{}_{;a}\,\lambda _c+h^b_c\,\left (Q^a\,\phi _{,a}
\right )_{,b}=0\ ,
\end{equation}
\begin{equation}\label{3.38}
\left (Q^a\,\phi _{,a}\right )\,\lambda _{b}=0\ .
\end{equation}
On account of (\ref{3.5}) the variable $y^a$ appearing in (\ref{3.31}), 
(\ref{3.34}) and (\ref{3.36}) is related to $x^a$ by
\begin{equation}\label{3.39}
y^a=\frac{dp}{d\mu}\,x^a\ .
\end{equation}
We must now examine the internal consistency of these equations.

Our first enquiry parallels the work in \cite{HE}. Putting
\begin{equation}\label{3.40}
V^{bt}=m^{bt}+i\,l^{bt}\ ,
\end{equation}
we can write (\ref{3.21}) and (\ref{3.24}) as a single complex 
equation
\begin{equation}\label{3.41}
2\,\dot\phi\,V^{bt}=i\,\eta ^{trsd}\,u_r\,V^b{}_s\,\phi _{,d}
+i\,\eta ^{brsd}\,u_r\,V^t{}_s\,\phi _{,d}\ .
\end{equation}
From this we calculate that
\begin{equation}\label{3.42}
\dot\phi\,\eta _{bpql}\,V^{bt}\,u^p=2\,i\,V^t{}_{[q}\,\lambda _{l]}\ .
\end{equation}
When this is substituted into each term on the right hand side of 
(\ref{3.41}) we obtain
\begin{equation}\label{3.43}
2\,\dot\phi\,V^{bt}=2\,\dot\phi\,V^{bt}+2\,\dot\phi ^{-1}\phi _{,d}\,
\phi ^{,d}\,V^{bt}\ .
\end{equation}
Hence with $\dot\phi\neq 0$ and $V^{bt}\neq 0$ we must have
\begin{equation}\label{3.44}
\phi _{,d}\,\phi ^{,d}=0\ .
\end{equation}
The hypersurfaces $\phi (x^a)={\rm constant}$ in the background 
isotropic cosmological model must be null (see section 4 below where 
the physical implications of (\ref{3.44}) are discussed). Thus $\lambda _a=h^b_a\,\phi _{,b}
=\phi _{,a}+\dot\phi\,u_a\neq 0$ and so we now have, in addition to 
(\ref{3.44}), the following vanishing scalar products:
\begin{equation}\label{3.45}
w^a\,\phi _{,a}=0\ ,\qquad a^a\,\phi _{,a}=0\ ,\qquad Q^a\,\phi _{,a}=0\ ,
\end{equation}
on account of (\ref{3.30}), (\ref{3.33}) and (\ref{3.38}). The latter 
two simplify (\ref{3.32}) and (\ref{3.37}) respectively.

We will next show that for consistency of our equations we must have $Q^a=0$. 
Let us, for convenience, write (\ref{3.8}) as 
\begin{equation}\label{3.46}
p^{ab}=A^{ab}-\dot s^{ab}-\frac{2}{3}\,\theta\,s^{ab}\ ,
\end{equation}
with
\begin{equation}\label{3.47}
A^{ab}=a^{(a;b)}+u^{(a}\,\dot a^{b)}-\frac{1}{3}\,\theta\,u^{(a}\,
a^{b)}-\frac{1}{3}\,a^f{}_{;f}\,h^{ab}\ .
\end{equation}
Using the fact that in the background $\theta _{,a}=-\dot\theta\,u_a$ we 
have from (\ref{3.46}),
\begin{equation}\label{3.48}
p^{ab}{}_{;b}=A^{ab}{}_{;b}-\dot s^{ab}{}_{;b}-\frac{2}{3}\,\theta\,
s^{ab}{}_{;b}\ .
\end{equation}
In the background $u_{a;b}=\frac{1}{3}\,\theta\,h_{ab}$ and so this 
equation can be written
\begin{equation}\label{3.49}
p^{ab}{}_{;b}=A^{ab}{}_{;b}-s^{ab}{}_{;cb}\,u^c-\theta\,s^{ab}{}_{;b}\ .
\end{equation}
The Ricci identities satisfied by $s^{ab}$ give
\begin{equation}\label{3.50}
s^{ab}{}_{;cb}\,u^c=\left (s^{ab}{}_{;b}\right )^.\ ,
\end{equation}
and so (\ref{3.49}) becomes
\begin{equation}\label{3.51}
p^{ab}{}_{;b}=A^{ab}{}_{;b}-\left (s^{ab}{}_{;b}\right )^.-\theta\,
s^{ab}{}_{;b}\ .
\end{equation}
Now (\ref{3.26}) is
\begin{equation}\label{3.52}
s^{ab}{}_{;b}=-Q^a+\frac{2}{3}\,z^a+{\cal A}^a\ ,
\end{equation}
with 
\begin{equation}\label{3.53}
{\cal A}^a=-\eta ^{acdf}\,u_c\,w_{d;f}\ .
\end{equation}
Putting (\ref{3.52}) into (\ref{3.51}) yields
\begin{equation}\label{3.54}
p^{ab}{}_{;b}=A^{ab}{}_{;b}+\dot Q^a+\theta\,Q^a-\frac{2}{3}\,\dot z^a-\frac{2}{3}\,\theta\,z^a-\dot 
{\cal A}^a-\theta\,{\cal A}^a\ .
\end{equation}
Alternatively from (\ref{3.15}) and (\ref{3.34}) we have
\begin{equation}\label{3.55}
p^{ab}{}_{;b}=\frac{1}{3}\,x^a-\frac{1}{3}\,\theta\,Q^a+y^a+\dot Q^a+(\mu +p)\,a^a\ .
\end{equation}
Thus (\ref{3.54}) and (\ref{3.55}) are consistent provided
\begin{equation}\label{3.56}
\frac{4}{3}\,\theta\,Q^a=\frac{2}{3}\,\left (\dot z^a+\theta\,z^a+\frac{1}{2}\,x^a+\frac{3}{2}\,
y^a\right )+(\mu +p)\,a^a+\dot {\cal A}^a+\theta\,{\cal A}-A^{ab}{}_{;b}
\ .\end{equation}
Making use of (\ref{3.28}) to write $w^a$ in terms of $a^a$ and using 
the propagation equation (\ref{3.27}) for $w^a$ along $u^a$ it follows 
that 
\begin{eqnarray}\label{3.57}
\dot {\cal A}^a+\theta\,{\cal A}^a&=& \frac{1}{2}\ddot a^a+\frac{1}{2}\,\theta\,\dot a^a-
\left\{\frac{1}{4}\,(\mu -p)-\frac{1}{6}\,\dot\theta-\frac{1}{9}\,\theta ^2\right\}\,
a^a\ \nonumber\\
&&-\frac{1}{2}\,h^{ac}\,\left (a^b{}_{;b}\right )_{,c}-\frac{1}{3}\,\theta\,a^b{}_{;b}\,u^a+\frac{1}{2}\,
a^{a;b}{}_{;b}\ .
\end{eqnarray}
Direct calculation from (\ref{3.47}) yields
\begin{eqnarray}\label{3.58}
A^{ab}{}_{;b}&=& \frac{1}{2}\ddot a^a+\frac{1}{2}\,\theta\,\dot a^a+
\left\{\frac{1}{4}\,(\mu -p)-\frac{1}{6}\,\dot\theta-\frac{2}{9}\,\theta ^2\right\}\,
a^a\ \nonumber\\
&&+\frac{1}{6}\,h^{ac}\,\left (a^b{}_{;b}\right )_{,c}-\frac{1}{3}\,
\theta\,a^b{}_{;b}\,u^a+\frac{1}{2}\,
a^{a;b}{}_{;b}\ .
\end{eqnarray}
Putting (\ref{3.57}) and (\ref{3.58}) into (\ref{3.56}) and using the 
background Raychaudhuri equation (\ref{3.3}) gives
\begin{equation}\label{3.59}
\frac{4}{3}\,\theta\,Q^a=\frac{2}{3}\,\left \{\dot z^a+\theta\,z^a+\frac{1}{2}\,x^a+\frac{3}{2}\,
y^a-h^{ac}\,\left (a^b{}_{;b}\right )_{,c}-\dot\theta\,a^a\right\}\ .
\end{equation}
It thus follows from (\ref{3.31}) that
\begin{equation}\label{3.60}
\theta\,Q^a=0\ .
\end{equation}
Since $\theta >0$ we must have
\begin{equation}\label{3.61}
Q^a=0\ .\end{equation}
It now follows from (\ref{3.37}) and (\ref{3.38}) that
\begin{equation}\label{3.62}
x^a=0\ ,
\end{equation}
and so (\ref{3.39}) yields
\begin{equation}\label{3.63}
y^a=0\ .
\end{equation}
Now (\ref{3.32})with (\ref{3.45}) becomes
\begin{equation}\label{3.64}
\dot\phi\,z^c=a^d{}_{;d}\,\lambda ^c\ ,
\end{equation}
while (\ref{3.36}) reduces to
\begin{equation}\label{3.65}
(\mu +p)\,(z^c+\theta\,a^c)=0\ .
\end{equation}
Contracting (\ref{3.64}) with $\phi _{,c}$ and using (\ref{3.44}) 
results in
\begin{equation}\label{3.66}
z^c\,\phi _{,c}=\dot\phi\,a^d{}_{;d}\ ,
\end{equation}
and, noting (\ref{3.45}) again, contracting (\ref{3.65}) with 
$\phi _{,c}$ gives 
\begin{equation}\label{3.67}
\dot\phi\,(\mu +p)\,a^d{}_{;d}=0\ .
\end{equation}
With $\dot\phi\neq 0\ ,\ \mu +p\neq 0$ we must have $a^d{}_{;d}=0$ 
and so (\ref{3.64}) becomes
\begin{equation}\label{3.68}
z^c=0\ .
\end{equation}
Now (\ref{3.65}) with $\theta >0$ and $\mu +p\neq 0$ yields
\begin{equation}\label{3.69}
a^c=0\ .
\end{equation}
It now follows from (\ref{3.28}) that
\begin{equation}\label{3.70}
w^a=0\ .\
\end{equation}

At this stage the only surviving gauge--invariant small quantities from 
the list (\ref{3.6}) are $\sigma ^{ab}$ and $\pi ^{ab}$ or equivalently 
$s^{ab}\ ,\ \Pi ^{ab}$. We see that now (\ref{3.25}) and (\ref{3.35}) 
require
\begin{equation}\label{3.71}
s^{ab}\,\phi _{,b}=0\ ,\qquad \Pi ^{ab}\,\phi _{,b}=0\ .
\end{equation}
Also (\ref{3.26}) and (\ref{3.34}) require $s^{ab}\ ,\ \Pi ^{ab}$ to 
satisfy
\begin{equation}\label{3.72}
s^{ab}{}_{;b}=0\ ,\qquad \Pi ^{ab}{}_{;b}=0\ .
\end{equation}
We will write the remaining equations from (\ref{3.13})--(\ref{3.38}), 
which have not reduced to $0=0$, in terms of $s^{ab}\ ,\Pi ^{ab}$. We will 
then check that the equations we obtain for $s^{ab}\ ,\ \Pi ^{ab}$ 
(including (\ref{3.71}) and (\ref{3.72})) are consistent. We 
begin with (\ref{3.20}) with $Q^a=0$ and substitute into it $p^{ab}$ 
from (\ref{3.8}) with $a^a=0$ and $m^{ab}$ from (\ref{3.9}) with $a^a=0$ 
to obtain
\begin{equation}\label{3.73}
\dot\phi\,\Pi ^{bt}=2\,\dot\phi\,\dot s^{bt}+\frac{5}{3}\,\theta\,
\dot\phi\,s^{bt}+\ddot\phi\,s^{bt}-u_r\,(q^{(b}{}_s\,\phi _{,d}+l^{(b}{}_{s;d})\,\eta ^{t)rsd}\ .
\end{equation}
With $q^{ab}$ given by (\ref{3.11}) with $w^a=0$ we calculate
\begin{equation}\label{3.74}
\eta ^{brsd}\,q^t{}_s\,u_r\,\phi _{,d}=-\dot s^{tp}\,\phi _{,p}\,u^b+
s^{tb;p}\,\phi _{,p}+\dot\phi\,\dot s^{tb}-s^{tp;b}\,\phi _{,p}+\frac{1}{3}\,\theta\,\dot\phi\,s^{tb}\ .
\end{equation}
Next using $l^{ab}$ in (\ref{3.12}) with $w^a=0$ we find
\begin{equation}\label{3.75}
\eta ^{brsd}\,u_r\,l_{st}=-\dot\phi\,s^d{}_t\,u^b+s^b{}_t\,\phi ^{,d}+\dot\phi\,s^b{}_t\,u^d-
s^d{}_t\,\phi ^{,b}\ .
\end{equation}
The first of (\ref{3.72}) and the background expression $u^d{}_{;d}=\theta$ help us 
to deduce from (\ref{3.75}) that
\begin{eqnarray}\label{3.76}
\eta ^{brsd}\,u_r\,l^t{}_{s;d}&=&-\dot\phi _{,d}\,s^{td}\,u^b-\frac{1}{3}\,
\theta\dot\phi\,s^{tb}+s^{tb}{}_{;d}\,\phi ^{,d}+s^{tb}\,\phi ^{,d}{}_{;d}
\nonumber\\
&&+\ddot\phi\,s^{tb}+\dot\phi\dot s^{tb}+\theta\,\dot\phi\,s^{tb}-\phi ^{,b}
{}_{;d}\,s^{td}\ .
\end{eqnarray}
Putting (\ref{3.74}) and (\ref{3.76}) together we get
\begin{equation}\label{3.77}
\eta ^{brsd}\,u_r\,(q^t{}_s\,\phi _{,d}+l^t{}_{s;d})=2\,s^{tb;d}\,\phi _{,d}
+\phi ^{,d}{}_{;d}\,s^{tb}+2\,\dot\phi\,\dot s^{tb}+\ddot\phi\,s^{tb}+\theta\,\dot\phi\,s^{tb}\ ,
\end{equation}
which is symmetric in $(t, b)$. Substituting this into (\ref{3.73}) gives 
the equation
\begin{equation}\label{3.78}
s'_{tb}+\left (\frac{1}{2}\,\phi ^{,d}{}_{;d}-\frac{1}{3}\,\theta\,\dot\phi
\right )\,s_{tb}=-\frac{1}{2}\,\dot\phi\,\Pi _{tb}\ ,
\end{equation}
with $s'_{tb}:=s_{tb;d}\,\phi ^{,d}$. {\it This is a propagation equation for 
$s_{tb}$ along the null geodesics tangent to $\phi ^{,d}$}.

We now turn our attention to (\ref{3.23}). Substituting for $p^{ab}\ ,\ 
m^{ab}$ from (\ref{3.8}) and (\ref{3.9}) with $a^a=0$ we see that
\begin{equation}\label{3.79}
p^b{}_s\,\phi _{,d}+m^b{}_{s;d}=-(\phi _{,d}\,s^b{}_s){}^.-\theta\,\phi 
_{,d}\,s^b{}_s-\frac{1}{3}\,\theta\,\dot\phi\,u_d\,s^b{}_s-\dot\phi\,s^b{}_{s;d}\ .
\end{equation}
We notice in passing from this that now (\ref{3.14}), which reads
\begin{equation}\label{3.80}
p^{ab}\,\phi _{,b}+m^{ab}{}_{;b}=0\ ,
\end{equation}
on account of the second of (\ref{3.71}), is automatically satisfied because 
of the first equation in (\ref{3.71}) and in (\ref{3.72}). Writing out 
(\ref{3.23}) with $q^{ab}$ substituted from (\ref{3.11}) with $w^a=0$ we 
have
\begin{equation}\label{3.81}
\dot l^{bt}+\theta\,l^{bt}-u_r\,(\dot\phi\,s^{(b}{}_{s;d}+p^{(b}{}_s\,\phi _{,d}
+m^{(b}{}_{s;d})\,\eta ^{t)rsd}=0\ .
\end{equation}
Now using (\ref{3.79}) we can write this as
\begin{equation}\label{3.82}
\dot l^{bt}+\theta\,l^{bt}+u_r\,\{(\phi _{,d}\,s^{(b}{}_s){}^.
+\theta\,\phi _{,d}\,s^{(b}{}_s\}\,\eta ^{t)rsd}=0\ .
\end{equation}
This can be re--arranged as 
\begin{equation}\label{3.83}
(l^{bt}+u_r\,\phi _{,d}\,s^{(b}{}_s\,\eta ^{t)rsd}){}^.+\theta\,
(l^{bt}+u_r\,\phi _{,d}\,s^{(b}{}_s\,\eta ^{t)rsd})=0\ .
\end{equation}
With $l^{bt}$ given in (\ref{3.12}) with $w^a=0$ we see that (\ref{3.83}) 
is identically satisfied. 

We next examine (\ref{3.22}). It is identically satisfied and this can be 
seen as follows: with $p^{ab}$ in (\ref{3.8}) and $a^a=0$, we find
\begin{equation}\label{3.84}
p^b{}_{s;d}=-\dot s^b{}_{s;d}+\frac{2}{3}\,\dot\theta\,u_d\,s^b{}_s-\frac{2}{3}\,
\theta\,s^b{}_{s;d}\ .
\end{equation}
From the Ricci identities satisfied by $s^{ab}$ we have
\begin{equation}\label{3.85}
\dot s^b{}_{s;d}=(s^b{}_{s;d}){}^.+(\frac{1}{6}\,\mu +\frac{1}{2}\,p)\,
(u^b\,s_{ds}+s^b{}_d\,u_s)
+\frac{1}{3}\,\theta\,(s^b{}_{s;d}+\dot s^b{}_s\,u_d)\ .
\end{equation}
This allows us to write
\begin{equation}\label{3.86}
\eta ^{trsd}\,u_r\,p^b{}_{s;d}=-(\eta ^{trsd}\,u_r\,s^b{}_{s;d}){}^.-
\theta\,\eta ^{trsd}\,u_r\,s^b{}_{s;d}\ .
\end{equation}
When this is entered into (\ref{3.22}) the equation can be re--arranged 
as
\begin{equation}\label{3.87}
(q^{bt}+u_r\,s^{(b}{}_{s;d}\,\eta ^{t)rsd}){}^.+
\theta\,(q^{bt}+u_r\,s^{(b}{}_{s;d}\,\eta ^{t)rsd})=0\ ,
\end{equation}
which is an identity on account of (\ref{3.11}) with $w^a=0$.

We now consider (\ref{3.19}) with $Q^a=0$. If we first substitute for 
$p^{ab}$ into it from (\ref{3.8}) with $a^a=0$ we obtain
\begin{equation}\label{3.88}
-\ddot s^{bt}-\frac{5}{3}\,\theta\,\dot s^{bt}-\frac{2}{3}\dot\theta\,s^{bt}-\frac{2}{3}\,\theta ^2s^{bt}+\frac{1}{2}
(\mu +p)\,s^{bt}+u_r\,q^{(b}{}_{s;d}\,\eta ^{t)rsd}=-\dot\Pi ^{bt}-\frac{2}{3}
\theta\,\Pi ^{bt}\ .
\end{equation}
However using $q^{ab}$ in (\ref{3.11}) with $w^a=0$ we find that
\begin{equation}\label{3.89}
\eta ^{brsd}\,u_r\,q^t{}_s=-\frac{1}{3}\,\theta\,s^{td}\,u^b+\frac{1}{3}\theta\,s^{tb}\,
u^d-\dot s^{td}\,u^b+\dot s^{bt}\,u^d-s^{td;b}+s^{tb;d}\ ,
\end{equation}
and thus, using the first of (\ref{3.72}),
\begin{equation}\label{3.90}
\eta ^{brsd}\,u_r\,q^t{}_{s;d}=\frac{2}{9}\,\theta ^2s^{tb}+\frac{1}{3}\,\dot\theta\,s^{tb}
+\ddot s^{tb}+\theta\,\dot s^{tb}-s^{td;b}{}_{;d}+s^{tb;d}{}_{;d}\ .
\end{equation}
The second last term here can be simplified using the Ricci identities satisfied 
by $s^{ab}$ and the first of (\ref{3.72}) to read
\begin{equation}\label{3.91}
s^{td;b}{}_{;d}=\left (\frac{5}{6}\,\mu -\frac{1}{2}\,p\right )\,s^{tb}\ .
\end{equation}
We see that now (\ref{3.90}) is symmetric in $(b, t)$ and on substitution 
into (\ref{3.88}) we arrive at {\it a wave equation for $s^{ab}$}, namely,
\begin{equation}\label{3.92}
s^{ab;d}{}_{;d}-\frac{2}{3}\,\theta\,\dot s^{ab}-\left (\frac{1}{3}\,\dot\theta 
+\frac{4}{9}\,\theta ^2\right )\,s^{ab}+(p-\frac{1}{3}\,\mu )\,
s^{ab}=-\dot\Pi ^{ab}-\frac{2}{3}\,\theta\,\Pi ^{ab}\ .
\end{equation}

With $w^a=0=Q^a$ we have from (\ref{3.18}) that
\begin{equation}\label{3.93}
q^{ab}{}_{;b}=0\ .
\end{equation}
This equation is satisfied by $q^{ab}$ given by (\ref{3.11}) with 
$w^a=0$. Substituting the latter expression for $q^{ab}$ into the 
left hand side of (\ref{3.93}) and using (\ref{3.91}) we have
\begin{eqnarray}\label{3.94}
q^{ab}{}_{;b}&=&-\frac{1}{4}\,(s^a{}_{p;cb}-s^a{}_{p;bc})\,\eta ^{bcfp}\,u_f
\ ,\nonumber\\
&=&-\frac{1}{4}\,\left (R^a{}_{gcb}\,s^g{}_p+R_{pgcb}\,s^{ag}\right )\,
\eta ^{bcfp}\,u_f\ ,
\end{eqnarray}
using the Bianchi identities. Here $R^a{}_{gcb}$ are the components of 
the Riemann tensor of the isotropic background. These components can be 
easily written in terms of the perfect--fluid energy--momentum--stress tensor 
of the background because the background, being isotropic, 
is conformally flat. When this is done it is found that each of the Riemann 
tensor terms in (\ref{3.94}) separately vanishes and so (\ref{3.93}) is 
satisfied.

With the second of (\ref{3.72}) holding we have from (\ref{3.15})
\begin{equation}\label{3.95}
p^{ab}{}_{;b}=0\ .
\end{equation}
That this is satisfied by $p^{ab}$ given by (\ref{3.8}) with $a^a=0$ is straightforward 
when one notes that $\dot s^b{}_{s;b}=(s^b{}_{;b}){}^.=0$, which follows 
from (\ref{3.85}) after summation over $(b, d)$ and the properties of $s^{ab}$. 

Finally we must check (\ref{3.17}) with $Q^a=0$. This reads
\begin{equation}\label{3.96}
q^d{}_b\,\phi _{,d}+l^d{}_{b;d}=0\ .
\end{equation}
To see that $q^{ab}$ and $l^{ab}$ satisfies this equation we start with (\ref{3.77}) and multiply it by $\eta _{bpqt}$. 
Since the right hand side of (\ref{3.77}) is symmetric in $(t, b)$ we obtain
\begin{equation}\label{3.97}
(q^m{}_p\,\phi _{,m}+l^m{}_{p;m})\,u_q-(q^m{}_q\,\phi _{,m}+l^m{}_{q;m})\,u_p=0
\ .
\end{equation}
If this is multiplied by $u^p$ then (\ref{3.96}) results.

At this point all of the equations (\ref{3.13})--(\ref{3.38}) are satisfied 
provided $s^{ab}\ ,\ \Pi ^{ab}$ satisfy the algebraic relations with $\phi _{,a}$ 
in (\ref{3.71}), are divergence--free as indicated in (\ref{3.72}) and satisfy 
the propagation equation (\ref{3.78}) for $s^{ab}$ along the integral curves of 
$\phi ^{,a}$ in the background space--time and satisfy the wave equation (\ref{3.92}) . 
These equations reduce to those obtained in \cite{HE} when $\Pi ^{ab}=0$. 
We need to check that (\ref{3.71}), (\ref{3.72}), (\ref{3.78}) and (\ref{3.92}) are 
consistent.

Using the Bianchi identities we can show that
\begin{eqnarray}\label{3.98}
(s^{ab;d}{}_{;d})_{;b}&=&(s^{ab}{}_{;b})^{;d}{}_{;d}+\left (\frac{7}{6}\,\mu 
-\frac{1}{2}\,p\right )\,
s^{ab}{}_{;b}\ ,\\
\dot s^{ab}{}_{;b}&=&(s^{ab}{}_{;b})^.+\frac{1}{3}\,\theta\,s^{ab}{}_{;b}\ ,\\
\dot \Pi ^{ab}{}_{;b}&=&(\Pi ^{ab}{}_{;b})^.+\frac{1}{3}\,\theta\,\Pi ^{ab}{}_{;b}\ .
\end{eqnarray}
With the help of these equations it is straightforward to see that the wave equation (\ref{3.92}) is 
consistent with (\ref{3.72}). Also using 
\begin{equation}\label{3.101}
s^{ab;d}{}_{;d}\,u_b=(s^{ab}\,u_b)^{;d}{}_{;d}-\frac{2}{3}\,\theta\,s^{ab}{}_{;b}\ ,
\end{equation}
one can easily see that the wave equation (\ref{3.92}) is consistent with 
$s^{ab}\,u_b=0=\Pi ^{ab}\,u_b$. The wave equation (\ref{3.92}) is also 
consistent with $s^{ab}\,\phi _{,b}=0=\Pi^{ab}\,\phi _{,b}$. This follows from 
\begin{equation}\label{3.102}
s^{ab;d}{}_{;d}\,\phi _{,b}=-2\,(s^{ab;d}\,\phi _{,d})_{;b}-s^{ab}\,
(\phi ^{,d}{}_{;d})_{,b}\ ,
\end{equation}
which is obtained using the Bianchi identities satisfied by $s^{ab}$ and $\phi _{,a}$. 
The propagation equation (\ref{3.78}) allows us to write (\ref{3.102}) as 
\begin{equation}\label{3.103}
s^{ab;d}{}_{;d}\,\phi _{,b}=-\frac{2}{3}\,\theta\,\dot\phi _{,b}\,s^{ab}+\dot\phi _{,b}
\,\Pi ^{ab}\ .
\end{equation}
One can also derive this by multiplying the wave equation (\ref{3.92}) by $\phi _{,b}$ 
and using the fact that $(\phi _{,b})^.=\dot\phi _{,b}-\frac{1}{3}\,\theta\,\lambda _b$ 
with, as always, $\lambda _b=h^c_b\,\phi _{,c}$.

The propagation equation (\ref{3.78}) is clearly consistent with (\ref{3.71}) because the 
integral curves of $\phi ^{,a}$ are geodesics. That (\ref{3.78}) is also consistent with 
$s^{ab}\,u_b=0=\Pi ^{ab}\,u_b$ follows from $u'_b=u_{b;c}\,\phi ^{,c}=\frac{1}{3}\,\theta\,\lambda _b$. 
The consistency of (\ref{3.78}) with (\ref{3.72}) requires the wave equation (\ref{3.92}). 
This is because on taking the divergence of (\ref{3.78}) and using the Ricci identities 
one arrives at
\begin{equation}\label{3.104}
-\frac{1}{2}\,\dot\phi\,\Pi ^{ab}{}_{;b}+\frac{1}{2}\,\phi _{,b}\,\dot\Pi ^{ab}=
-\frac{1}{2}\phi _{,b}\,s^{ab;d}{}_{;d}+\frac{1}{3}\,\theta\,\dot s^{ab}\,\phi _{,b}\ .
\end{equation}
Substituting for $s^{ab;d}{}_{;d}$ here from the wave equation (\ref{3.92}), this equation 
reduces to $\Pi ^{ab}{}_{;b}=0$.

\setcounter{equation}{0}
\section{Pure Gravity Wave Perturbations}\indent

As a result of the calculations outlined in section 3 the perturbations 
of the Weyl tensor now have ``electric" and ``magnetic" parts given by
\begin{eqnarray}\label{4.1}
E_{ab}&=&\left (\frac{1}{2}\,\Pi _{ab}+p_{ab}\right )F+m_{ab}\,F'\ ,\\
H_{ab}&=&q_{ab}\,F+l_{ab}\,F'\ ,
\end{eqnarray}
with
\begin{eqnarray}\label{4.3}
p_{ab}&=&-\dot s_{ab}-\frac{2}{3}\,\theta\,s_{ab}\ ,
\qquad m_{ab}=-\dot\phi\,s_{ab}\ ,\\
q_{ab}&=&-s_{(a}{}^{p;c}\,\eta _{b)fpc}\,u^f\ ,\qquad 
l_{ab}=-s_{(a}{}^p\,\eta _{b)fpc}\,u^f\,\phi ^{,c}\ .
\end{eqnarray}
Also $s^{ab}\ ,\ \Pi ^{ab}$ satisfy the consistent 
equations (\ref{3.71}), (\ref{3.72}), (\ref{3.78}) and (\ref{3.92}) 
and $\phi ^{,a}$ is a null vector field in the background RW 
space--time. From the first of (\ref{3.71}) we see from (\ref{4.3}) 
and (4.4) that
\begin{equation}\label{4.5}
m^{ab}\,\phi _{,b}=0\ ,\qquad l^{ab}\,\phi _{,b}=0\ ,
\end{equation}
verifying that (\ref{3.13}) and (\ref{3.16}) are satisfied. Thus the 
$F'$--parts of $E_{ab}\ ,\ H_{ab}$ above are type N in the Petrov 
classification with degenerate principal null 
direction $\phi ^{,a}$. We therefore consider the $F'$--part of this 
perturbed field as describing gravitational waves having propagation direction $\phi ^{,a}$ 
in the RW background and the histories of the wave--fronts are the 
null hypersurfaces $\phi (x^a)={\rm constant}$. This interpretation is 
based on the well--known analogy with electromagnetic radiation \cite{Pirani}. 
The $F$--parts of 
$E_{ab}\ ,\ H_{ab}$ are not in general type N and so do not necessarily 
describe gravitational waves.

For the remainder of this paper we will consider pure type N 
perturbations (i.e. pure gravity wave perturbations) of the RW background. 
We could do this by requiring the $F$--parts of $E_{ab}\ ,\ H_{ab}$ in 
(\ref{4.1}) and (4.2) to vanish. It is possible to exhibit solutions 
of our basic equations (3.71), (3.72), (3.78) and (3.92) having this 
property (see section 6 below) but it is less restrictive to require that the 
$F$--parts of $E_{ab}\ ,\ H_{ab}$ also be type N in the Petrov classification 
with $\phi ^{,a}$ as degenerate principal null direction. This means that, 
in the light of the second of (\ref{3.71}), we should require
\begin{equation}\label{4.6}
p^{ab}\,\phi _{,b}=0\ ,\qquad q^{ab}\,\phi _{,b}=0\ ,
\end{equation}
with $p^{ab}\ ,\ q^{ab}$ given above in (\ref{4.3}) and (4.4). The 
first of these can be written 
\begin{equation}\label{4.7}
s^{ab}\,\phi _{,b;c}\,u^c=0\ ,
\end{equation}
while the second gives us
\begin{equation}\label{4.8}
s^{ab}\,\phi _{,a}{}^{;c}-s^{ac}\,\phi _{,a}{}^{;b}=0\ .
\end{equation}
To elucidate the meaning of (\ref{4.7}) and (\ref{4.8}) it 
is convenient to make use of a null tetrad in the background 
RW space--time. First we note that $k_a=-\dot\phi ^{-1}\phi _{,a}$ 
and $l_a=u_a-\frac{1}{2}\,k_a$ are two real covariant vector fields 
satisfying $k_a\,k^a=0 ,\ l_a\,l^a=0$ and $k_a\,l^a=-1$. Let $m_a ,\ 
\bar m_a$ be a complex covariant vector field and its complex 
conjugate (indicated by a bar) chosen so that they are null ($m_a\,m^a
=0=\bar m_a\,\bar m^a$), are orthogonal to $k^a$ and $l^a$ and 
satisfy $m_a\,\bar m^a=1$. Now $k^a ,\ l^a ,\ m^a ,\ \bar m^a$ 
constitute a null tetrad with respect to which $s^{ab}$ can be 
written (because $s^a{}_a=0\ ,\ s^{ab}\,u_b=0\ ,\ s^{ab}\,k_b=0$ 
and so $s^{ab}\,l_b=0$)
\begin{equation}\label{4.9}
s^{ab}=\bar s\,m^a\,m^b+s\,\bar m^a\,\bar m^b\ .
\end{equation}
Thus $|s|^2=\frac{1}{2}\,s^{ab}\,s_{ab}$. Substituting (\ref{4.9}) 
into (\ref{4.7}) we easily see that (\ref{4.7}) is equivalent to 
\begin{equation}\label{4.10}
s\,\phi _{,b;c}\,\bar m^b\,l^c=0\ ,
\end{equation}
from which we conclude that provided $s\neq 0\,(\Leftrightarrow 
s^{ab}\neq 0$) we must have 
\begin{equation}\label{4.11}
\phi _{,b;c}\,\bar m^b\,l^c=0\ .
\end{equation}
On using (\ref{4.9}) in (\ref{4.8}) we find that, in addition to (\ref{4.11}),
\begin{equation}\label{4.12}
\bar s\,\phi _{,a;b}\,m^a\,m^b=s\,\phi _{,a;b}\,\bar m^a\,\bar m^b\ .
\end{equation}
A simple way to satisfy this with $s\neq 0$ is to require the null 
hypersurfaces $\phi (x^a)={\rm constant}$ to satisfy
\begin{equation}\label{4.13}
\phi _{,a;b}\,m^a\,m^b=0\ .
\end{equation} 
This means that {\it the complex shear of the 
null geodesic congruence tangent to $\phi ^{,a}$ in the background 
RW space--time vanishes}. If we can find a family of null hypersurfaces 
$\phi (x^a)={\rm constant}$ in the background space--time satisfying 
(\ref{4.11}) and (\ref{4.13}) then (\ref{4.6}) will be satisfied. The 
solutions we then obtain of the equations (3.71), (3.72), (3.78) and 
(3.92) will be analogous to the Bateman waves \cite{B} of electromagnetic theory.

\setcounter{equation}{0}
\section{Explicit Examples}\indent

To exhibit explicit examples of pure type N perturbations of the RW 
space--times we first select in such space--times some naturally 
occuring shear--free null hypersurfaces $\phi (x^a)={\rm constant}$. We 
begin with the general Robertson--Walker line--element in 
standard form,
\begin{equation}\label{5.1}
ds^2=R^2(t)\,\frac{[(dx^1)^2+(dx^2)^2+(dx^3)^2]}{\left (1+\frac{k}{4}\,r^2\right )^2}
-dt^2\ ,
\end{equation}
where $R(t)$ is the scale factor, $r^2=(x^1)^2+(x^2)^2+(x^3)^2$ and 
$k=0\ ,\ \pm 1$ is the Gaussian curvature of the space--like hypersurfaces 
$t={\rm constant}$. We can put these line--elements in the following 
interesting forms for our purposes \cite{H}:
\begin{equation}\label{5.2}
ds^2=R^2(t)\,\{dx^2+p_0^{-2}f^2(dy^2+dz^2)\}-dt^2\ ,
\end{equation}
with $p_0=1+\frac{K}{4}\,(y^2+z^2)\ ,\ K={\rm constant}\ ,\ f=f(x)$. The 
following cases arise: (i) if $k=+1$ then $K=+1$ and $f(x)=\sin x$; 
(ii) if $k=0$ then $K=0, +1$ with $f(x)=1$ when $K=0$ and $f(x)=x$ 
when $K=+1$; (iii) if $k=-1$ then $K=0, \pm 1$ with $f(x)=\frac{1}{2}\,
{\rm e}^x$ when $K=0$, $f(x)=\sinh x$ when $K=+1$ and $f(x)=\cosh x$ 
when $K=-1$. 

Case (i) above arises because when $k=+1$ the closed model universe with 
line--element (\ref{5.1}) has $t={\rm constant}$ hypersurfaces with line--element 
which can be put in the form $dl^2=R^2(t)\,ds_0^2$ with 
\begin{equation}\label{5.3}
ds_0^2=dx^2+\sin ^2x\,(d\vartheta ^2+\sin ^2\vartheta\,d\varphi ^2)\ ,
\end{equation}
and we then use stereographic coordinates $y, z$ such that $y+iz=2\,{\rm e}^{i\varphi}\,
\cot\frac{\vartheta}{2}$ in place of the polar angles $\vartheta\ ,\ \varphi$.

Case (ii) arises because when $k=0$ the open, spatially flat universe with 
line--element (\ref{5.1}) has $t={\rm constant}$ hypersurfaces with 
line--element $dl^2=R^2(t)\,ds_0^2$ where
\begin{equation}\label{5.4}
ds_0^2=dx^2+dy^2+dz^2\ ,
\end{equation}
or 
\begin{equation}\label{5.5}
ds_0^2=dx^2+x^2(d\vartheta ^2+\sin ^2\vartheta\,d\varphi ^2)\ ,
\end{equation}
and in the latter we introduce the stereographic coordinates $y, z$ again 
in place of $\vartheta\ ,\ \varphi$.

Case (iii) is due to the fact that in (\ref{5.1}) when $k=-1$ the 
$t={\rm constant}$ hypersurfaces can, modulo the factor $R^2(t)$, 
each be viewed as the future sheet of a unit time--like hypersphere 
${\cal H}_3$ in four dimensional Minkowskian space--time ${\cal M}_4$. Thus 
if the line--element of ${\cal M}_4$ is written
\begin{equation}\label{5.6}
ds_0^2=(dz^1)^2+(dz^2)^2+(dz^3)^2-(dz^4)^2\ ,
\end{equation}
then ${\cal H}_3$ is given by
\begin{equation}\label{5.7}
(z^1)^2+(z^2)^2+(z^3)^2-(z^4)^2=-1\ ,\qquad  z^4>0\ .
\end{equation}
The different parts of case (iii) above are due to the different ways one can 
parametrise (\ref{5.7}). One possibility is with 
\begin{eqnarray}\label{5.8}
z^1+iz^2&=&(y+iz)\,p_0^{-1}\sinh x\ ,\\
z^3&=&\left (\frac{1}{4}(y^2+z^2)-1\right )\,p_0^{-1}\sinh x\ ,\\
z^4&=&\cosh x\ ,
\end{eqnarray}
with $p_0=1+\frac{1}{4}(y^2+z^2)$. Substitution into (\ref{5.6}) gives
\begin{equation}\label{5.11}
ds_0^2=dx^2+p_0^{-2}\sinh ^2x\,(dy^2+dz^2)\ .
\end{equation}
The next possibility is 
\begin{eqnarray}\label{5.12}
z^1+iz^2&=&\frac{1}{2}\,{\rm e}^x\,(y+iz)\ ,\\
z^3&=&\frac{1}{4}\,{\rm e}^x\,(y^2+z^2-1)+{\rm e}^{-x}\ ,\\
z^4&=&\frac{1}{4}\,{\rm e}^x\,(y^2+z^2+1)+{\rm e}^{-x}\ ,
\end{eqnarray}
and now (\ref{5.6}) reads
\begin{equation}\label{5.15}
ds_0^2=dx^2+\frac{1}{4}\,{\rm e}^{2\,x}\,(dy^2+dz^2)\ .
\end{equation}
Finally we can take
\begin{eqnarray}\label{5.16}
z^1+iz^2&=&(y+iz)\,p_0^{-1}\cosh x\ ,\\
z^3&=&\sinh x\ ,\\
z^4&=&\left (\frac{1}{4}\,(y^2+z^2)+1\right )\,p_0^{-1}\cosh x\ ,
\end{eqnarray}
with $p_0=1-\frac{1}{4}\,(y^2+z^2)$. With this (\ref{5.6}) takes the 
form
\begin{equation}\label{5.19}
ds_0^2=dx^2+p_0^{-2}\cosh ^2x\,(dy^2+dz^2)\ .
\end{equation}

In the space--times with line--elements (\ref{5.2}), with the 
special cases outlined following (\ref{5.2}), the hypersurfaces 
\begin{equation}\label{5.20}
\phi (x^a):=x-T(t)={\rm constant}\ ,
\end{equation}
with $dT/dt=R^{-1}$ are {\it null hypersurfaces}. They are generated 
by null geodesics having expansion
\begin{equation}\label{5.21}
\frac{1}{2}\,\phi ^{,a}{}_{;a}=\frac{f'}{R^2f}+\frac{\dot R}{R^2}\ .
\end{equation}
Here $f'=df/dx\ ,\ \dot R=dR/dt$. The integral curves of the vector field $\partial/
\partial t$ are the world--lines of the fluid particles. The components of 
this vector field are denoted by $u^a$ and using (\ref{5.20}) we 
can show that
\begin{equation}\label{5.22}
2\,\phi _{,a;b}=\xi _a\,\phi _{,b}+\xi _b\,\phi _{,a}+
\phi _{,d}{}^{;d}\,g_{ab}\ ,
\end{equation}
with
\begin{equation}\label{5.23}
\xi _a=-\frac{f'}{f}\,\phi _{,a}+R\,\phi _{,d}{}^{;d}\,u_a\ .
\end{equation}
With $s^{ab}\,\phi _{,b}=0=s^{ab}\,u_b$ we see on substituting 
(\ref{5.22}) that (\ref{4.7}) and (\ref{4.8}) are now satisfied. On 
account of (\ref{5.22}) it follows that {\it $\phi _{,a}$ is shear--free} 
\cite{RT}. Alternatively we can easily verify (\ref{4.11}) and (\ref{4.13}) 
using the null tetrad described following (\ref{4.8}) which is given 
via the 1--forms
\begin{eqnarray}\label{5.24}
k_a\,dx^a&=&R\,dx-dt\ ,\qquad l_a\,dx^a=-\frac{1}{2}\,(R\,dx+dt)\ ,
\nonumber\\
m_a\,dx^a&=&\frac{1}{\sqrt{2}}R\,p_0^{-1}f\,(dy+i\,dz)\ .
\end{eqnarray}

For convenience we have used the same coordinate labels $\{x, y, z, t\}$ 
for all of the special cases included in (\ref{5.2}). Of course the 
ranges of some of these coordinates will be different in the different cases 
(for example, in case (ii) $x\in (-\infty\ ,\ +\infty )$ if $K=0$ whereas 
$x\in [0, +\infty )$ if $K=+1$). Similarly the shear--free null 
hypersurfaces (\ref{5.20}) differ from case to case, and within 
cases (ii) and (iii), as can be seen by noting the intersections 
of these null hypersurfaces with the space--like hypersurfaces 
$t={\rm constant}$. In case (i) the intersection is a 2--sphere. In 
case (ii) it is a 2--sphere if $K=+1$ and a 2--plane if $K=0$. Thus 
(\ref{5.20}) describes two quite different families of shear--free 
null hypersurfaces that can arise in an open, spatially flat universe. 
In case (iii) the intersection of (\ref{5.20}) with $t={\rm constant}$ 
can be a 2--space of positive ($K=+1$), negative ($K=-1$) or 
zero ($K=0$) curvature giving three different families of shear--free 
null hypersurfaces in a $k=-1$ open universe. A geometrical explanation 
for these subcases is given in \cite{H}.

We begin with $\phi (x^a)$ given by (\ref{5.20}) and $u^a\partial /\partial x^a=
\partial /\partial t$. Since $s^{ab}$ and $\Pi ^{ab}$ are 
orthogonal to $u^a$ and $\phi ^{,a}$ and trace--free, 
with respect to the metric tensor given via the line--element (\ref{5.2}), 
each have only two independent components. If the coordinates are labelled 
$x^1=x\ , x^2=y\ , x^3=z\ , x^4=t$ then the surviving components are 
$s^{33}=-s^{22}=\alpha (x, y, z, t)\ ,\ s^{23}=s^{32}=\beta (x, y, z, t)$ 
and $\Pi ^{33}=-\Pi ^{22}=A(x, y, z, t)\ ,\ \Pi ^{23}=\Pi ^{32}=B(x, y, z, t)$. We 
can conveniently express these on the null tetrad (\ref{5.24}). We have 
$s^{ab}$ given by (\ref{4.9}) with 
\begin{equation}\label{5.25}
\bar s=-R^2p_0^{-2}f^2(\alpha +i\,\beta )\ ,
\end{equation}
and 
\begin{equation}\label{5.26}
\Pi ^{ab}=\bar\Pi\,m^a\,m^b+\Pi\,\bar m^a\,\bar m^b\ ,
\end{equation}
with
\begin{equation}\label{5.27}
\bar\Pi =-R^2p_0^{-2}f^2(A+i\,B)\ .
\end{equation}
Calculation of the first of (\ref{3.72}) to be satisfied by $s^{ab}$ shows 
that $\alpha\ ,\ \beta$ must satisfy the Cauchy--Riemann equations
\begin{equation}\label{5.28}
\frac{\partial}{\partial y}(p_0^{-4}\alpha )-\frac{\partial}{\partial z}
(p_0^{-4}\beta )=0\ ,
\end{equation}
\begin{equation}\label{5.29}
\frac{\partial}{\partial y}(p_0^{-4}\beta )+\frac{\partial}{\partial z}
(p_0^{-4}\alpha )=0\ .
\end{equation}
In addition $\Pi ^{ab}$ and thus $A\ ,\ B$ must satisfy the same equations,
\begin{equation}\label{5.30}
\frac{\partial}{\partial y}(p_0^{-4}A )-\frac{\partial}{\partial z}
(p_0^{-4}B )=0\ ,
\end{equation}
\begin{equation}\label{5.31}
\frac{\partial}{\partial y}(p_0^{-4}B )+\frac{\partial}{\partial z}
(p_0^{-4}A )=0\ .
\end{equation}
We shall find it convenient to work with $\alpha _0\ ,\ \beta _0$ rather 
than $\alpha\ ,\ \beta$ where
\begin{equation}\label{5.32}
\alpha _0=\alpha\,f^3R^3\ ,\qquad \beta _0=\beta\,f^3R^3\ .
\end{equation}
Since $f=f(x)\ ,\ R=R(t)$ we have (\ref{5.28}) and (\ref{5.29}) 
satisfied by $\alpha _0$ and $\beta _0$ and these equations can be 
written economically as
\begin{equation}\label{5.33}
\frac{\partial}{\partial\bar\zeta}\{p_0^{-4}(\alpha _0+i\,\beta _0 )\}=0\ ,
\end{equation}
with $\zeta =y+i\,z$, giving
\begin{equation}\label{5.34}
\alpha _0+i\,\beta _0=p_0^4\,{\cal G}(\zeta , x, t )\ .
\end{equation}
where ${\cal G}$ is an analytic function of $\zeta$. Now (\ref{5.25}) 
reads
\begin{equation}\label{5.35}
\bar s=-R^{-1}p_0^2f^{-1}{\cal G}(\zeta , x, t )\ .
\end{equation}
The propagation equation (\ref{3.78}) for $s^{ab}$ along the integral 
curves of $\phi ^{,a}$ gives $A , B$ in terms of $\alpha _0 , \beta _0$. 
Writing this in terms of $\Pi$ given by (\ref{5.27}) we find that 
\begin{equation}\label{5.36}
\bar\Pi =-2\,R^{-2}p_0^2f^{-1}(D{\cal G}+\dot R\,{\cal G})\ .
\end{equation}
Here the operator $D$ is given by $D=\partial /\partial x+R\,\partial /
\partial t= \partial /\partial x+\partial /\partial T$ with $T(t)$ 
introduced in (\ref{5.20}). Also the dot indicates differentiation of $R(t)$. 
It follows from this and (\ref{5.27}) that $A+i\,B$ is analytic in $\zeta$ 
and so (\ref{5.30}) and (\ref{5.31}) are automatically satisfied. The 
only remaining equation to satisfy is the wave equation (\ref{3.92}). With 
$s^{ab}$ given by (\ref{4.9}) and (\ref{5.35}) and with $\Pi ^{ab}$ 
given by (\ref{5.26}) and (\ref{5.36}) we find, after a lengthy 
calculation, that (\ref{3.92}) reduces to the remarkably simple wave 
equation
\begin{equation}\label{5.37}
D^2{\cal G}+k\,{\cal G}=0\ ,
\end{equation}
with $k=0, \pm 1$ labelling the RW backgrounds with line--elements of 
the form (\ref{5.2}). 
Thus we have for $k=0$, 
\begin{equation}\label{5.38}
{\cal G}(\zeta , x, t )=a(\zeta , x-T)\,(x+T)+b(\zeta , x-T)\ ,
\end{equation}
for $k=+1$, 
\begin{equation}\label{5.39}
{\cal G}(\zeta , x, t )=a(\zeta , x-T)\,\sin\left (\frac{x+T}{2}\right )+b(\zeta , x-T)\,
\cos\left (\frac{x+T}{2}\right )\ ,
\end{equation}
and for $k=-1$, 
\begin{equation}\label{5.40}
{\cal G}(\zeta , x, t )=a(\zeta , x-T)\,\sinh\left (\frac{x+T}{2}\right )+b(\zeta , x-T)\,
\cosh\left (\frac{x+T}{2}\right )\ ,
\end{equation}
where in each case $a(\zeta , x-T) ,\ b(\zeta , x-T)$ are arbitrary functions. In 
deriving (\ref{5.37}) we have made use of the equations
\begin{equation}\label{5.41}
f''=-k\,f\ ,\qquad (f')^2+k\,f^2=K\ ,
\end{equation}
which are satisfied in the cases (i)--(iii) described following (\ref{5.2}) above. 

With $s^{ab}$ and $\Pi ^{ab}$ known we can calculate $m^{ab}\ ,\ l^{ab}\ ,\ 
p^{ab}$ and $q^{ab}$ in order to form the electric and magnetic parts of the 
perturbed Weyl tensor as indicated in (\ref{4.1}) and (4.2). We can 
write the result compactly as
\begin{equation}\label{5.42}
E^{ab}+i\,H^{ab}=-2\,R^{-2}p_0^2f^{-1}\frac{\partial}{\partial x}
({\cal G}\,F)\,m^a\,m^b\ .
\end{equation}
We emphasise that ${\cal G}$ is given by (\ref{5.38})--(\ref{5.40}) in 
the various cases and now $F=F(x-T)$ so that $F'=\partial F/\partial x$. 
Also $p_0=1+\frac{K}{4}(y^2+z^2)\ ,\ f=f(x)$ described following (\ref{5.2}), 
and $R(t)$ is the scale factor. It is immediately clear from (\ref{5.42}) 
that the perturbations of the RW background which we have constructed 
here are pure gravity wave perturbations. We will discuss some of their 
properties in section 6.
\vfill\eject

\setcounter{equation}{0}
\section{Discussion}\indent

The propagation equation (\ref{3.78}) for $s^{ab}$ along the null 
geodesics tangent to $\phi ^{,a}$ shows that if $s^{ab}=0$ then 
$\Pi ^{ab}=0$. An important converse property of the pure type N 
perturbations described in section 5 is that {\it if $\Pi ^{ab}=0$ 
then $s^{ab}=0$ provided $\mu +p\neq 0$}. To see this we have from 
(\ref{5.36}) that $\Pi ^{ab}=0$ is equivalent to 
\begin{equation}\label{6.1}
D{\cal G}+\dot R\,{\cal G}=0\ .
\end{equation}
Substituting this into the wave equation (\ref{5.37}) results in 
\begin{equation}\label{6.2}
(\dot R^2-R\,\ddot R+k)\,{\cal G}=0\ .
\end{equation}
For the background RW space--time the fluid proper density $\mu$ and 
isotropic pressure $p$ satisfy
\begin{equation}\label{6.3}
\frac{2}{R^2}\,(\dot R^2-R\,\ddot R+k)=\mu +p\ ,
\end{equation}
as a consequence of Einstein's field equations. Hence we can rewrite 
(\ref{6.2}) simply as
\begin{equation}\label{6.4}
(\mu +p)\,{\cal G}=0\ .
\end{equation}
From this and (\ref{5.35}) it follows that $s^{ab}=0$ provided 
$\mu +p\neq 0$.

The perturbed Weyl tensor given via (\ref{5.42}) for the pure type N 
perturbations can be infinite where $p_0 (y, z)$ is infinite (when 
$y^2+z^2\rightarrow +\infty$ if $K\neq 0$) and where $f(x)$ vanishes 
(when $k=+1$ at $x=0$, when $k=-1$ at $x\rightarrow -\infty\ (K=0)$ 
or at $x=0\ (K=0)$). {\it There is one non--singular case 
corresponding to $k=0\ ,\ K=+1$} for which $p_0 =1$ and $f=1$. In this 
case the expansion of the history of the wave--fronts (\ref{5.21}) 
is entirely due to the expansion of the universe. This case is as 
close as one can get to plane--waves in the present context and is 
analogous to plane Bateman waves in electromagnetic theory.

Had we wished to construct examples of type N perturbations 
for which the $F$--parts of (\ref{4.1}) and (4.2) vanish 
we see from (\ref{5.42}) that these would be given by (\ref{5.42}) 
with $\partial {\cal G}/\partial x=0$. This condition would then 
be incorporated into the wave equation (\ref{5.37}) and the 
appropriate solutions ${\cal G}(\zeta , t)$ replacing (\ref{5.38})--
(\ref{5.40}) could easily be obtained.

We note that we have used the assumption $\theta >0$ in the background 
cosmological models to conclude from (3.60) that $Q^a=0$. If the background 
were an Einstein static universe then $\theta =0$ and we would have 
$Q^a\neq 0$. It is well--known (see, for example \cite{Peebles}) that 
the Einstein universe is unstable and it might be interesting to 
investigate how this instability manifests itself in the formalism we 
use in this paper.

There are exact cosmological solutions of Einstein's field equations known 
which contain gravitational waves (see \cite{Wain}, \cite{BG} and references 
therein). These solutions describe universes with a stiff equation of state so 
that the speed of sound is equal to the speed of light. Our perturbations 
describing gravitational waves propagating through isotropic cosmologies place 
no restriction on the equation of state of the isotropic background and 
thus we would not in general expect them to approximate to these known 
exact solutions.

\noindent
\section*{Acknowledgment}\noindent

We thank Professor Englebert Sch\"ucking for suggesting that viscosity 
should play a role in the effect of gravity waves on the matter content 
of the universe. The important part played by anisotropic stress in 
this paper confirms this point of view. One of us (E.O'S) wishes to 
thank Enterprise Ireland for the award of a Postgraduate Scholarship.

\vskip 8truepc

\end{document}